\documentclass[notoc]{JHEP3}
\usepackage{graphicx}
\usepackage{amsmath,amsfonts}
\usepackage{color}
\let\normalcolor\relax

\usepackage{lscape}
\usepackage{multirow}

\def\hermesauthor[#1]#2{{#2}$^{\, #1}$}
\def\hermesinstitute[#1]#2{$^{#1\,}$ {#2}\\}

\def\nowat[#1]#2{\(^,\)\footnote[#1]{#2}}

\def\desy{{\sc Desy}}

\title{
 	Exclusive Leptoproduction of Real Photons on \\
        a Longitudinally Polarised Hydrogen Target
}

\author{The HERMES Collaboration}

\abstract{
Polarisation asymmetries are measured for the hard exclusive leptoproduction of
real photons from a longitudinally polarised hydrogen target. These
asymmetries arise from the deeply virtual
Compton scattering and Bethe-Heitler processes. From the data are
extracted two asymmetries in the azimuthal distribution of produced real photons
about the direction of the exchanged virtual photon: $\mathcal{A}_{\text{UL}}$
with respect to the target polarisation and $\mathcal{A}_{\text{LL}}$ with
respect to the product of the beam and target polarisations. Results for both
asymmetries are compared to the predictions from a generalised
parton distribution model. The $\sin\phi$ and $\cos(0\phi)$
amplitudes observed respectively for the $\mathcal{A}_{\text{UL}}$ and
$\mathcal{A}_{\text{LL}}$ asymmetries are compatible with the sizeable
predictions from the model. Unexpectedly, a $\sin(2\phi)$ modulation in the
$\mathcal{A}_{\text{UL}}$ asymmetry with a magnitude similar to that of the
$\sin\phi$ modulation is observed.
}

\begin{document}
%\collab{HERMES Collaboration}
{
\bf
\begin{flushleft}
\hermesauthor[12,15]{A.~Airapetian},
\hermesauthor[26]{N.~Akopov},
\hermesauthor[5]{Z.~Akopov},
\hermesauthor[6]{E.C.~Aschenauer}\nowat[1]{Now at: Brookhaven National Laboratory, Upton, New York 11772-5000, USA},
\hermesauthor[25]{W.~Augustyniak},
\hermesauthor[26]{R.~Avakian},
\hermesauthor[26]{A.~Avetissian},
\hermesauthor[5]{E.~Avetisyan},
\hermesauthor[15]{B.~Ball},
\hermesauthor[18]{S.~Belostotski},
\hermesauthor[10]{N.~Bianchi},
\hermesauthor[17,24]{H.P.~Blok},
\hermesauthor[6]{H.~B\"ottcher},
\hermesauthor[5]{A.~Borissov},
\hermesauthor[13]{J.~Bowles},
\hermesauthor[12]{I.~Brodski},
\hermesauthor[19]{V.~Bryzgalov},
\hermesauthor[13]{J.~Burns},
\hermesauthor[9]{M.~Capiluppi},
\hermesauthor[10]{G.P.~Capitani},
\hermesauthor[21]{E.~Cisbani},
\hermesauthor[9]{G.~Ciullo},
\hermesauthor[9]{M.~Contalbrigo},
\hermesauthor[9]{P.F.~Dalpiaz},
\hermesauthor[5]{W.~Deconinck}\nowat[2]{Now at: Massachusetts Institute of Technology, Cambridge, Massachusetts 02139, USA},
\hermesauthor[2]{R.~De~Leo},
\hermesauthor[11,5]{L.~De~Nardo},
\hermesauthor[10]{E.~De~Sanctis},
\hermesauthor[14,8]{M.~Diefenthaler},
\hermesauthor[10]{P.~Di~Nezza},
\hermesauthor[12]{M.~D\"uren},
\hermesauthor[12]{M.~Ehrenfried},
\hermesauthor[26]{G.~Elbakian},
\hermesauthor[4]{F.~Ellinghaus}\nowat[3]{Now at: Institut f\"ur Physik, Universit\"at Mainz, 55128 Mainz, Germany},
\hermesauthor[6]{R.~Fabbri},
\hermesauthor[10]{A.~Fantoni},
\hermesauthor[22]{L.~Felawka},
\hermesauthor[21]{S.~Frullani},
\hermesauthor[11,6]{D.~Gabbert},
\hermesauthor[19]{G.~Gapienko},
\hermesauthor[19]{V.~Gapienko},
\hermesauthor[21]{F.~Garibaldi},
\hermesauthor[5,18,22]{G.~Gavrilov},
\hermesauthor[26]{V.~Gharibyan},
\hermesauthor[5,9]{F.~Giordano},
\hermesauthor[15]{S.~Gliske},
\hermesauthor[6]{M.~Golembiovskaya},
\hermesauthor[10]{C.~Hadjidakis},
\hermesauthor[5]{M.~Hartig}\nowat[4]{Now at: Institut f\"ur Kernphysik, Universit\"at Frankfurt a.M., 60438 Frankfurt a.M., Germany},
\hermesauthor[10]{D.~Hasch},
\hermesauthor[13]{G.~Hill},
\hermesauthor[6]{A.~Hillenbrand},
\hermesauthor[13]{M.~Hoek},
\hermesauthor[5]{Y.~Holler},
\hermesauthor[6]{I.~Hristova},
\hermesauthor[23]{Y.~Imazu},
\hermesauthor[19]{A.~Ivanilov},
\hermesauthor[18]{A.~Izotov},
\hermesauthor[1]{H.E.~Jackson},
\hermesauthor[11]{H.S.~Jo},
\hermesauthor[14,11]{S.~Joosten},
\hermesauthor[13]{R.~Kaiser},
\hermesauthor[26]{G.~Karyan},
\hermesauthor[13,12]{T.~Keri},
\hermesauthor[4]{E.~Kinney},
\hermesauthor[18]{A.~Kisselev},
\hermesauthor[23]{N.~Kobayashi},
\hermesauthor[19]{V.~Korotkov},
\hermesauthor[16]{V.~Kozlov},
\hermesauthor[18]{P.~Kravchenko},
\hermesauthor[7]{V.G.~Krivokhijine},
\hermesauthor[2]{L.~Lagamba},
\hermesauthor[14]{R.~Lamb},
\hermesauthor[17]{L.~Lapik\'as},
\hermesauthor[13]{I.~Lehmann},
\hermesauthor[9]{P.~Lenisa},
\hermesauthor[14]{L.A.~Linden-Levy},
\hermesauthor[11]{A.~L\'opez~Ruiz},
\hermesauthor[15]{W.~Lorenzon},
\hermesauthor[6]{X.-G.~Lu},
\hermesauthor[23]{X.-R.~Lu},
\hermesauthor[3]{B.-Q.~Ma},
\hermesauthor[13]{D.~Mahon},
\hermesauthor[14]{N.C.R.~Makins},
\hermesauthor[18]{S.I.~Manaenkov},
\hermesauthor[21]{L.~Manfr\'e},
\hermesauthor[3]{Y.~Mao},
\hermesauthor[25]{B.~Marianski},
\hermesauthor[4]{A.~Martinez de la Ossa},
\hermesauthor[26]{H.~Marukyan},
\hermesauthor[22]{C.A.~Miller},
\hermesauthor[23]{Y.~Miyachi},
\hermesauthor[26]{A.~Movsisyan},
\hermesauthor[10]{V.~Muccifora},
\hermesauthor[13]{M.~Murray},
\hermesauthor[5,8]{A.~Mussgiller},
\hermesauthor[2]{E.~Nappi},
\hermesauthor[18]{Y.~Naryshkin},
\hermesauthor[8]{A.~Nass},
\hermesauthor[6]{M.~Negodaev},
\hermesauthor[6]{W.-D.~Nowak},
\hermesauthor[9]{L.L.~Pappalardo},
\hermesauthor[12]{R.~Perez-Benito},
\hermesauthor[8]{N.~Pickert},
\hermesauthor[8]{M.~Raithel},
\hermesauthor[1]{P.E.~Reimer},
\hermesauthor[10]{A.R.~Reolon},
\hermesauthor[6]{C.~Riedl},
\hermesauthor[8]{K.~Rith},
\hermesauthor[13]{G.~Rosner},
\hermesauthor[5]{A.~Rostomyan},
\hermesauthor[14]{J.~Rubin},
\hermesauthor[11]{D.~Ryckbosch},
\hermesauthor[19]{Y.~Salomatin},
\hermesauthor[20]{F.~Sanftl},
\hermesauthor[20]{A.~Sch\"afer},
\hermesauthor[6,11]{G.~Schnell},
\hermesauthor[5]{K.P.~Sch\"uler},
\hermesauthor[13]{B.~Seitz},
\hermesauthor[23]{T.-A.~Shibata},
\hermesauthor[7]{V.~Shutov},
\hermesauthor[9]{M.~Stancari},
\hermesauthor[9]{M.~Statera},
\hermesauthor[8]{E.~Steffens},
\hermesauthor[17]{J.J.M.~Steijger},
\hermesauthor[12]{H.~Stenzel},
\hermesauthor[6]{J.~Stewart},
\hermesauthor[8]{F.~Stinzing},
\hermesauthor[26]{S.~Taroian},
\hermesauthor[16]{A.~Terkulov},
\hermesauthor[25]{A.~Trzcinski},
\hermesauthor[11]{M.~Tytgat},
\hermesauthor[11]{A.~Vandenbroucke},
\hermesauthor[17]{P.B.~van~der~Nat},
\hermesauthor[11]{Y.~Van~Haarlem}\nowat[5]{Now at: Carnegie Mellon University, Pittsburgh, Pennsylvania 15213, USA},
\hermesauthor[11]{C.~Van~Hulse},
\hermesauthor[18]{D.~Veretennikov},
\hermesauthor[18]{V.~Vikhrov},
\hermesauthor[2]{I.~Vilardi},
\hermesauthor[8]{C.~Vogel},
\hermesauthor[3]{S.~Wang},
\hermesauthor[6,8]{S.~Yaschenko},
\hermesauthor[3]{H.~Ye},
\hermesauthor[5]{Z.~Ye},
\hermesauthor[22]{S.~Yen},
\hermesauthor[12]{W.~Yu},
\hermesauthor[8]{D.~Zeiler},
\hermesauthor[5]{B.~Zihlmann},
\hermesauthor[25]{P.~Zupranski}
\end{flushleft}
}
\bigskip
{\it
\begin{flushleft}
\hermesinstitute[1]{Physics Division, Argonne National Laboratory, Argonne, Illinois 60439-4843, USA}
\hermesinstitute[2]{Istituto Nazionale di Fisica Nucleare, Sezione di Bari, 70124 Bari, Italy}
\hermesinstitute[3]{School of Physics, Peking University, Beijing 100871, China}
\hermesinstitute[4]{Nuclear Physics Laboratory, University of Colorado, Boulder, Colorado 80309-0390, USA}
\hermesinstitute[5]{DESY, 22603 Hamburg, Germany}
\hermesinstitute[6]{DESY, 15738 Zeuthen, Germany}
\hermesinstitute[7]{Joint Institute for Nuclear Research, 141980 Dubna, Russia}
\hermesinstitute[8]{Physikalisches Institut, Universit\"at Erlangen-N\"urnberg, 91058 Erlangen, Germany}
\hermesinstitute[9]{Istituto Nazionale di Fisica Nucleare, Sezione di Ferrara and Dipartimento di Fisica, Universit\`a di Ferrara, 44100 Ferrara, Italy}
\hermesinstitute[10]{Istituto Nazionale di Fisica Nucleare, Laboratori Nazionali di Frascati, 00044 Frascati, Italy}
\hermesinstitute[11]{Department of Subatomic and Radiation Physics, University of Gent, 9000 Gent, Belgium}
\hermesinstitute[12]{Physikalisches Institut, Universit\"at Gie{\ss}en, 35392 Gie{\ss}en, Germany}
\hermesinstitute[13]{Department of Physics and Astronomy, University of Glasgow, Glasgow G12 8QQ, United Kingdom}
\hermesinstitute[14]{Department of Physics, University of Illinois, Urbana, Illinois 61801-3080, USA}
\hermesinstitute[15]{Randall Laboratory of Physics, University of Michigan, Ann Arbor, Michigan 48109-1040, USA }
\hermesinstitute[16]{Lebedev Physical Institute, 117924 Moscow, Russia}
\hermesinstitute[17]{National Institute for Subatomic Physics (Nikhef), 1009 DB Amsterdam, The Netherlands}
\hermesinstitute[18]{Petersburg Nuclear Physics Institute, Gatchina, Leningrad district 188300, Russia}
\hermesinstitute[19]{Institute for High Energy Physics, Protvino, Moscow region 142281, Russia}
\hermesinstitute[20]{Institut f\"ur Theoretische Physik, Universit\"at Regensburg, 93040 Regensburg, Germany}
\hermesinstitute[21]{Istituto Nazionale di Fisica Nucleare, Sezione Roma 1, Gruppo Sanit\`a and Physics Laboratory, Istituto Superiore di Sanit\`a, 00161 Roma, Italy}
\hermesinstitute[22]{TRIUMF, Vancouver, British Columbia V6T 2A3, Canada}
\hermesinstitute[23]{Department of Physics, Tokyo Institute of Technology, Tokyo 152, Japan}
\hermesinstitute[24]{Department of Physics \& Astronomy, VU University, 1081 HV Amsterdam, The Netherlands}
\hermesinstitute[25]{Andrzej Soltan Institute for Nuclear Studies, 00-689 Warsaw, Poland}
\hermesinstitute[26]{Yerevan Physics Institute, 375036 Yerevan, Armenia}
\end{flushleft}
}

\clearpage

\section{Introduction}
Generalised Parton Distributions (GPDs)~\cite{D.Mueller1994,Ji1997,Radyushkin1996} encompass the familiar parton distribution functions and form factors within a unified description of the nucleon. The study of the concept of GPDs has shown that they may provide a way to investigate the contribution of quark orbital angular momentum to the spin of the nucleon~\cite{Ji1997}. In a frame in which the proton moves quickly in the ``longitudinal'' direction, GPDs contain correlated information on the monodimensional distribution of quark momentum fractions in that direction with their two-dimensional spatial distributions in the transverse plane~\cite{PhysRevD.62.071503}. Access to GPDs can be achieved through the measurement of cross-sections and asymmetries in the exclusive production of photons and mesons leaving an intact nucleon~\cite{Diehl2003}. There are four chiral-even quark GPDs for the nucleon at leading twist: $H$, $E$, $\widetilde{H}$, and $\widetilde{E}$. The $H$ and $E$ distributions are quark helicity-averaged distributions, whereas the $\widetilde{H}$ and $\widetilde{E}$ distributions involve quark helicity differences. The $H$ and $\widetilde{H}$ distributions conserve nucleon helicity, while the $E$ and $\widetilde{E}$ distributions are associated with a change in nucleon helicity. 

Appearing in the descriptions of exclusive leptoproduction of photons or mesons, GPDs are dependent on four kinematic variables:~$x$, $\xi$, $t$, and $Q^2$. The variables $x$ and $\xi$ are the average and half the difference of the longitudinal parton momenta as fractions of the ``infinite'' nucleon momentum in the initial and final states respectively, in a frame in which the initial proton moves quickly.
The average fraction $x$ is not directly experimentally accessible and the skewness variable $\xi$ is related to the Bjorken scaling variable $x_{\textrm{B}}=\frac{-q^2}{2p\cdot q}$ as $\xi \simeq \frac{x_{\textrm{B}}}{(2-x_{\textrm{B}})}$ in the Bjorken limit where  the virtuality of the exchanged photon $Q^2\equiv\,-q^2\rightarrow\infty$, while $x_\text{B}$ and $t$ are fixed. Here, $p$ is the four-momentum of the target nucleon and $q$ is the four-momentum of the virtual photon. There is currently no consensus as to how to define $\xi$ precisely in terms of experimental observables at finite $Q^2$; the results are reported as projections in $x_{\text{B}}$. The Mandelstam $t$ variable is defined as the squared momentum transfer to the target nucleon, \textit{i.e.} $t=(p-p')^2$ where $p'$ is the four-momentum of the recoiling nucleon.  The dependence of GPDs on  $Q^2$ is implicit because GPDs are subject to quantum chromodynamics evolution with $Q^2$, which has been calculated perturbatively to leading order and next-to-leading order in the strong coupling constant $\alpha_s$~\cite{D.Mueller1994,Ji1997,Radyushkin1996,Bluemlein1999,Belitsky2000}.

The Deeply Virtual Compton Scattering (DVCS) process ($l\,p\,\rightarrow l\,\,\gamma\,p$) is the simplest of those currently measurable that provide access to GPD-related information. At leading-order, a quark in the nucleon absorbs the virtual photon $\gamma^{\ast}$ (which mediates the electromagnetic interaction between the incident lepton $l$ and the target nucleon $p$) radiated by the incident lepton, and itself radiates a detectable real photon. This combination of initial ($l\,p$) and final ($l\,\gamma\,p$) states appears in two experimentally indistinguishable processes, the DVCS and Bethe-Heitler (BH) processes. Unlike in the DVCS process, in the BH process the real photon is radiated from the initial or the scattered lepton.

In this paper, asymmetries in the distribution of photons produced from a hydrogen target by a positron beam are measured, where both the beam and target are polarised parallel or antiparallel to the beam direction. The extracted asymmetry amplitudes are compared to those predicted by a GPD model~\cite{Vanderhaeghen1999}.

The four-fold differential cross-section for exclusive leptoproduction of real photons, neglecting target polarisation components transverse to the direction of the virtual photon, is given as 
\begin{equation}\label{eqn:xsec}
\frac{ \text{d} \sigma }{ \text{d}x_{\text{B}} \,\text{d}Q^{2}\,
\text{d}\lvert t\rvert \,\text{d} \phi } = \frac{ x_{\text{B}} \,e^{6}\, \lvert\tau\rvert^2} {32 (2\pi)^{4} \,Q^{4}\,
\sqrt{ 1+\epsilon^{2} }
}  \,,
\end{equation}
where $e$ is the elementary charge, the angle $\phi$ is defined as the azimuthal angle between the
plane containing the directions of the incident and scattered lepton trajectories and the plane containing the trajectories of the virtual and real photons~\cite{Bacchetta2004}, and $\epsilon\equiv2x_{\textrm{B}}\frac{M}{Q}$, in which $M$ is the mass
of the nucleon. The square of the scattering amplitude $\lvert\tau\rvert^{2}$ can be written as
\begin{equation}\label{eqn:interfere}\lvert\tau\lvert^{2} = \lvert
\tau_{\text{BH}}\rvert^{2} + \lvert \tau_{\text{DVCS}}\rvert^{2}	+
\overbrace{\tau_{\text{BH}}\tau^{\ast}_{\text{DVCS}} +
\tau_{\text{BH}}^{\ast}\tau_{\text{DVCS}}}^{\textrm{I}}\,.
\end{equation}
Although the squared-DVCS amplitude $\lvert\tau_{\text{DVCS}}\rvert^2$ is small relative to the squared-BH amplitude $\lvert\tau_{\text{BH}}\rvert^2$ in the kinematic range
of the HERMES experiment~\cite{Korotkov2002}, information on GPDs can be
accessed via the Interference term denoted $\textrm{I}$, which arises from the interference of the scattering amplitudes of the two processes~\cite{Belitsky2002}.
To leading order in the electromagnetic coupling constant $\alpha_{em}$, the Fourier expansion of these quantities for a charged lepton beam and a target nucleon, where both the beam and the target are longitudinally polarised, reads:

\begin{eqnarray}\label{eqn:BHscatXsec}
\vert\tau_{\text{BH}}\vert^{2} & = &  \frac{K_{\text{BH}}}{\mathcal{P}_1(\phi)\mathcal{P}_2(\phi)} \left\lbrace
\sum\limits^{2}_{n=0} c_{n,\text{unp}}^{\text{BH}}\cos(n\phi) + P_{z}\, P_{\ell} \sum\limits^{1}_{n=0} c_{n,\text{LP}}^{\text{BH}}\cos(n\phi)  \right\rbrace\,, \\
\label{eqn:DVCSscatXsec}
\vert\tau_{\text{DVCS}}\vert^{2} & = &
K_{\text{DVCS}}\Bigg\lbrace \sum\limits^{2}_{n=0} c_{n,\text{unp}}^{\text{DVCS}}\cos(n\phi) +  P_{\ell}\,s_{1,\text{unp}}^{\text{DVCS}}\sin\phi \nonumber\\ & & +   P_{z}\left[ P_{\ell} \sum\limits^{1}_{n=0} c_{n,\text{LP}}^{\text{DVCS}}\cos(n\phi) + \sum\limits^{2}_{n=1} s_{n,\text{LP}}^{\text{DVCS}}\sin(n\phi) \right]     \Bigg\rbrace\,, \\
\label{eqn:IscatXsec}
\text{I} & = &
\frac{-e_{\ell}\,K_{\text{I}}}{\mathcal{P}_1(\phi)\mathcal{P}_2(\phi)
}\Bigg\lbrace \sum\limits^3_{n=0}c_{n,\text{unp}}^{\text{I}}\cos(n\phi) + P_{\ell}\sum\limits^2_{n=1} s_{n,\text{unp}}^{\text{I}}\sin(n\phi) \nonumber\\ & & +  P_{z}\left[ P_{\ell} \sum\limits^{2}_{n=0} c_{n,\text{LP}}^{\text{I}}\cos(n\phi) + \sum\limits^{3}_{n=1} s_{n,\text{LP}}^{\text{I}}\sin(n\phi) \right]   \Bigg\rbrace\,.
\end{eqnarray}
Here $\mathcal{P}_{1}(\phi)$ and $\mathcal{P}_{2}(\phi)$ are lepton propagators of the BH process, $P_{\ell}$ and $e_{\ell}$ respectively represent the longitudinal polarisation and the charge of the lepton
beam in units of the elementary charge and $P_z$ represents the longitudinal polarisation of the target. The subscript $\textrm{unp}$ ($\textrm{LP}$) denotes coefficients for an unpolarised (longitudinally polarised) target. The terms $K_{\text{BH}}$, $K_{\text{DVCS}}$ and $K_\textrm{I}$ are kinematic factors: 
$K_{\text{BH}}=1/\left(x_\textrm{B}^2\,t\,\left(1+\epsilon^2\right)^2\right)$,
 $K_{\text{DVCS}}=1/Q^2$ and $K_{\textrm{I}}=1/\left(x_{\textrm{B}}\,y\,t\right)$, where $y$ is the fraction of the beam energy carried by the virtual photon in the target rest frame. The lepton propagators and BH coefficients $c^{\text{BH}}_{n,\text{(unp$\vert$LP)}}$ can be calculated in QED with the latter also having a dependence on $F_1$ and $F_2$, the
Dirac and Pauli form
factors of the nucleon. The Fourier coefficients $c^{\text{DVCS}}_{n,\text{(unp$\vert$LP)}}$ and $s^{\text{DVCS}}_{n,\text{(unp$\vert$LP)}}$ arising from the squared-DVCS term relate to a bilinear combination of GPDs, whereas the Fourier coefficients $c^{\text{I}}_{n,\text{(unp$\vert$LP)}}$ and $s^{\text{I}}_{n,\text{(unp$\vert$LP)}}$ from the Interference term relate to a linear combination of GPDs.

Two asymmetries in the azimuthal distribution of real photons for positron scattering from a longitudinally polarised proton target are
presented in this paper: one single-spin asymmetry $\mathcal{A}_{\text{UL}}$ arising from the longitudinal
polarisation of the
target averaged over all beam polarisation states, and one double-spin asymmetry
$\mathcal{A}_{\text{LL}}$ arising from the longitudinal polarisation of both beam and target. These can be written as

\begin{align}
\mathcal{A}_{\text{UL}}(\phi,e_{\ell})  &\equiv \frac{\left[\sigma^{\leftarrow\Rightarrow}(\phi,e_{\ell})+
\sigma^{\rightarrow\Rightarrow}(\phi,e_{\ell})\right] - \left[\sigma^{\leftarrow\Leftarrow}(\phi,e_{\ell}) +
\sigma^{\rightarrow\Leftarrow}(\phi,e_{\ell})\right]  }{\left[\sigma^{\leftarrow\Rightarrow}(\phi,e_{\ell})+
\sigma^{\rightarrow\Rightarrow}(\phi,e_{\ell})\right] + \left[\sigma^{\leftarrow\Leftarrow}(\phi,e_{\ell}) +
\sigma^{\rightarrow\Leftarrow}(\phi,e_{\ell})\right] } \label{eqn:AUL_defn}\\
&=  \frac{  K_{\text{DVCS}}
\displaystyle\sum_{n=1}^{2}s_{n,\textrm{LP}}^{\text{DVCS}}\sin(n\phi) -
\textstyle\frac{e_{\ell}K_{\textrm{I}}}{\mathcal{P}_{1}(\phi)\mathcal{P}_{2}
(\phi)} \displaystyle\sum_{n=1}^{3}s_{n,\textrm{LP}}^{\textrm{I}}\sin(n\phi)
}{\frac{1}{\mathcal{P}_{1}(\phi)\mathcal{P}_{2}(\phi)} \Biggl[ \textstyle
K_{\text{BH}}\displaystyle\sum_{n=0}^{2}c_{n,\textrm{unp}}^{\text{BH}}\cos(n\phi) -
\textstyle
e_{\ell}K_{\textrm{I}}\displaystyle\sum_{n=0}^{3}c_{n,\textrm{unp}}^{\textrm{I}}
\cos(n\phi) \Biggr]  + K_{\text{DVCS}}
\displaystyle\sum_{n=0}^{2}c_{n,\textrm{unp}}^{\text{DVCS}} \cos(n\phi) }\label{eqn:AUL_effective}\\
\mathcal{A}_{\text{LL}}(\phi,e_{\ell}) &\equiv
\frac{\left[\sigma^{\rightarrow\Rightarrow}(\phi,e_{\ell})+ 
\sigma^{\leftarrow\Leftarrow}(\phi,e_{\ell})\right] - \left[\sigma^{\leftarrow\Rightarrow}(\phi,e_{\ell}) +
\sigma^{\rightarrow\Leftarrow}(\phi,e_{\ell})\right]  }{ 
\left[\sigma^{\rightarrow\Rightarrow}(\phi,e_{\ell})+ \sigma^{\leftarrow\Leftarrow}(\phi,e_{\ell})\right] +
\left[\sigma^{\leftarrow\Rightarrow}(\phi,e_{\ell}) + \sigma^{\rightarrow\Leftarrow}(\phi,e_{\ell})\right] }\,,
\label{eqn:ALL_defn}\\  
&=      \frac{ \frac{\textstyle K_{\text{BH}}}{\mathcal{P}_{1}(\phi)\mathcal{P}_{2}(\phi)}\displaystyle\sum_{n=0}^{1}
c_{n,\textrm{LP}}^{\text{BH}}\cos(n\phi) + \textstyle
K_{\text{DVCS}}\displaystyle\sum_{n=0}^{1}
c_{n,\textrm{LP}}^{\text{DVCS}}\cos(n\phi)  -
\textstyle\frac{e_{\ell}K_{\textrm{I}}}{\mathcal{P}_{1}(\phi)\mathcal{P}_{2}
(\phi)} \displaystyle\sum_{n=0}^{2} c_{n,\textrm{LP}}^{\textrm{I}}\cos(n\phi)  }{
\frac{1}{\mathcal{P}_{1}(\phi)\mathcal{P}_{2}(\phi)} \Biggl[ \textstyle
K_{\text{BH}}\displaystyle\sum_{n=0}^{2}c_{n,\textrm{unp}}^{\text{BH}}\cos(n\phi) -
\textstyle
e_{\ell}K_{\textrm{I}}\displaystyle\sum_{n=0}^{3}c_{n,\textrm{unp}}^{\textrm{I}}
\cos(n\phi) \Biggr]  + \textstyle K_{\text{DVCS}}
\displaystyle\sum_{n=0}^{2}c_{n,\textrm{unp}}^{\text{DVCS}}\cos(n\phi)}\label{eqn:ALL_effective},
\end{align}
where $\sigma$ denotes the cross-section from Eq.~\ref{eqn:xsec}, $\rightarrow
(\leftarrow)$ represents the 
beam helicity state parallel (anti-parallel) to the beam momentum, and $\Leftarrow
(\Rightarrow)$ represents the target polarisation state parallel (anti-parallel) to the beam momentum. In the case of positron scattering, $e_{\ell}=+1$.

\section{Experiment and Event Selection}
\label{sec:ees}
Data were collected in 1996 and 1997 with the
HERMES spectrometer~\cite{Ackerstaff1998a} using a longitudinally polarised 27.6 GeV positron beam
incident on a longitudinally polarised hydrogen gas target~\cite{HERMESCollaboration2005} in the HERA lepton storage ring at DESY.  The integrated luminosity of the data sample analysed 
in this paper is approximately 50\,pb$^{-1}$~\cite{Benisch2001}. The average beam and target polarisations for the data are presented in Table~\ref{tab:Beam-Target}. 
\TABLE[t]{
\begin{tabular}[c]{|c||c|c|c|c|c|}
\cline{1-6} \multirow{2}{*}{Year} & Luminosity & \multicolumn{2}{|c|}{Beam Polarisation} & \multicolumn{2}{|c|}{Target Polarisation} \\
 & [pb$^{-1}$] & $P_{\ell}\,<\,0$ & $P_{\ell}\,>\,0$ & $P_{z}\,<\,0$ & $P_{z}\,>\,0$ \\ \cline{1-6}
\hline
1996 & 12.6 $\pm$ 1.0 &$-$ & 0.514 $\pm$ 0.017 & $-$0.759 $\pm$ 0.042  & 0.759 $\pm$ 0.042 \\ \cline{1-6}
1997 & 37.3 $\pm$ 3.2 & $-$0.531 $\pm$ 0.018 & 0.497 $\pm$ 0.017& $-$0.850 $\pm$ 0.032& 0.850 $\pm$ 0.032 \\ \cline{1-6}
\end{tabular}
\caption{The integrated luminosities in pb$^{-1}$ of the data sets, with average beam and target polarisations.}
\label{tab:Beam-Target}
}
A brief description of the event selection is given in the following paragraphs. More details can be found in Refs.~\cite{Airapetian2007b,Kopytin,Mahon2010}.
Exactly one charged track identified as a positron and one photon were
required within the acceptance of the spectrometer.  The kinematic
requirements imposed on each event are $1\,\textrm{GeV}^{2}<Q^{2}<10\,\textrm{GeV}^{2}$, $0.03<x_{\text{B}}<0.35$, $W>3\,\textrm{GeV}$, and $\nu<22\,\textrm{GeV}$, where $W$ is the invariant mass of the initial $\gamma^{\ast}p$ state and $\nu$ is the energy of the virtual photon in the target rest frame as determined from measurements of the energies of the positron before and after scattering.

Photons were identified by selecting a single signal cluster in the electromagnetic calorimeter with no associated track in the rest of the spectrometer. The cluster was required to register an energy deposition of at least $5\,\text{GeV}$ in the calorimeter in order to reduce background, and to register at least $1\,\text{MeV}$ in the preshower detector in order to improve the resolution of the energy measurement.

The polar angle between the virtual and real photons was required to be between $5\,\textrm{mrad}$ and $45\,\textrm{mrad}$. The lower limit on this requirement was imposed in order to ensure that the azimuthal angle $\phi$ remains well-defined within the finite resolution of the spectrometer -- it excludes very few events from the data sample. The upper limit on the requirement was determined by Monte Carlo (MC) studies~\cite{Ellinghaus2004}, which indicated that the data set above this value is dominated by background from the Bethe-Heitler process producing resonant states of the proton and from semi-inclusive meson production. 

As the recoiling proton was not detected, events were selected by requiring that the squared missing-mass
$M_{\text{X}}^{2}=({q}+{p}-{q'})^{2}$ of the $e\,p\rightarrow e\,\gamma\,\text{X}$ reaction corresponded to the squared proton
mass within the spectrometer resolution, where $q'$ is the four-momentum of the real photon.  An ``exclusive region'' in the missing-mass distribution was determined from MC simulations described in Ref.~\cite{Airapetian2008}.  Corrections were applied to account for shifts in the $M_{\text{X}}^{2}$ distributions between data samples for different years.

The value of $t$ was determined from the kinematics of the scattered lepton and the real photon under the assumption of $ep\rightarrow ep\gamma$ in order to avoid using the measurement of the energy of the real photon, which was subject to the largest experimental uncertainty. The requirement $-t<0.7\,\textrm{GeV}^{2}$ was imposed in the exclusive region to further reduce background contamination~\cite{Ellinghaus2004}.

\section{Experimental Extraction of Asymmetry Amplitudes}
This paper presents Fourier amplitudes of the asymmetries of Eqs.~\ref{eqn:AUL_defn} and~\ref{eqn:ALL_defn}, rather than the related Fourier coefficients of the cross-section, which appear in Eqs.~\ref{eqn:BHscatXsec}--\ref{eqn:IscatXsec}. The link between them is clarified in the following section.

Section~\ref{sec:ees} describes the selection of an event yield $\mathcal{N}$. Its expectation value can be written as
\begin{equation}\label{eqn:yield}
\left\langle\mathcal{N}\left(P_{\ell},\,P_{z},\,\phi, e_{\ell}\right)\right\rangle =  \mathcal{L}\left(P_{\ell}\right)\,\eta(\phi)\,\sigma_{\text{UU}}(\phi, e_{\ell})\left[1+P_{z}\,\mathcal{A}_{\text{UL}}(\phi) + P_{\ell}\,P_{z}\,\mathcal{A}_{\text{LL}}(\phi) + P_{\ell}\,\mathcal{A}_{\text{LU}}(\phi)\right]\,,
\end{equation}
where $\mathcal{L}$ is the integrated luminosity, $\eta$ the detection efficiency and $\sigma_{\text{UU}}$ denotes the cross-section for an unpolarised beam  and an unpolarised target. The beam helicity asymmetry $\mathcal{A}_{\text{LU}}$ is not considered in this paper since the dataset presented here is a subset of data previously analysed with respect to the beam helicity~\cite{Airapetian2009}. In analogy to the decomposition of the cross-section in Eqs.~\ref{eqn:BHscatXsec}--\ref{eqn:IscatXsec}, the asymmetries defined in Eqs.~\ref{eqn:AUL_defn} and \ref{eqn:ALL_defn} can be decomposed as
\begin{center}
\begin{equation}
\label{eqn:AUL_fit}
\mathcal{A}_{\text{UL}}(\phi) \simeq
\sum^{3}_{n=1} A_{\text{UL}}^{\sin(n\phi)} \sin(n\phi) +
A_{\text{UL}}^{\cos(0\phi)}\,,
\end{equation}
\begin{equation}
\label{eqn:ALL_fit}\mathcal{A}_{\text{LL}}(\phi) \simeq
\sum^{2}_{n=0} A_{\text{LL}}^{\cos(n\phi)}
\cos(n\phi)\,.
\end{equation}
\end{center}
The azimuthal asymmetry amplitudes $A$ in Eqs.~\ref{eqn:AUL_fit} and~\ref{eqn:ALL_fit} are extracted from the data using the maximum likelihood fitting method~\cite{Airapetian2008,Barlow1990}, with each amplitude containing a combination of the Fourier coefficients from Eqs.~\ref{eqn:BHscatXsec}--\ref{eqn:IscatXsec} with the exception of $A_{\text{UL}}^{\cos(0\phi)}$. This term has no physical meaning. It is included only as a test of the normalisation of the function and is expected to be zero. 

Previously published measurements with transversely polarised~\cite{Airapetian2008} and unpolarised targets~\cite{Airapetian2009,Airapetian2009b} were made with both electron and positron beams. The beam-charge dependence of the contribution from the Interference term to the cross-section then allowed the separation of the squared-DVCS and Interference terms via charge difference and charge average asymmetries. However, the longitudinally polarised hydrogen data set at HERMES was taken solely with a positron beam, so the separation of squared-DVCS and Interference terms is not possible.

The asymmetry amplitudes in Eqs.~\ref{eqn:AUL_fit} and~\ref{eqn:ALL_fit} are distinguished mainly by the Fourier coefficients in the numerators of Eqs.~\ref{eqn:AUL_effective} and~\ref{eqn:ALL_effective} (see the first two columns in Table~\ref{tab:coefficients}). The extracted amplitudes may also be influenced by the $\phi$-dependent lepton propagators and/or the other $\phi$-dependent terms in the denominators of Eqs.~\ref{eqn:AUL_effective} and~\ref{eqn:ALL_effective}. 

Correlations between the asymmetry amplitudes determined by the fit were found to be small, and likelihood ratio tests for higher-order terms show a null result. Therefore, only the terms shown in Eqs.~\ref{eqn:AUL_fit} and~\ref{eqn:ALL_fit} were fitted to the data, and the result of that fit is presented in this paper.

The correspondences between the individual asymmetry amplitudes from Eqs.~\ref{eqn:AUL_fit}
and \ref{eqn:ALL_fit} and the Fourier coefficients in the decomposition of the differential cross-section (within a kinematic factor) from Eqs.~\ref{eqn:BHscatXsec}--\ref{eqn:IscatXsec}, which are interpretable within the GPD framework, are
clarified in Table \ref{tab:coefficients}. The relation of these Fourier 
coefficients to GPDs is encompassed in $\mathcal{C}$-functions~\cite{Belitsky2002,Diehl1997,Ji1998,Belitsky2001}. These functions depend upon combinations of Compton Form Factors (CFFs), which are convolutions of GPDs with hard scattering kernels, and consequently have real and imaginary parts. Furthermore, the CFF information contained in the $\mathcal{C}$-functions enters with various degrees of suppression. Higher twist terms are, in general, suppressed by powers of $\frac{1}{Q}$ compared to leading twist (twist-2). There are two $\mathcal{C}$-functions that appear at twist-2 in the observables reported in this paper, $\mathcal{C}_{\text{LP}}^{\text{I}}$ and $\mathcal{C}_{\text{LP}}^{\text{DVCS}}$. While the latter is a bilinear combination of CFFs and their complex conjugates, the former is written
\begin{center}
\begin{equation}
\label{eqn:twist-2_C_LP}
\mathcal{C}_{\textrm{LP}}^{\textrm{I}}  =  
\frac{x_{\textrm{B}}}{2-x_{\textrm{B}}} \left(F_{1}+F_{2}\right)\left(\mathcal{H} +
\frac{x_{\textrm{B}}}{2}\mathcal{E} \right) + F_{1}\widetilde{\mathcal{H}} -
\frac{x_{\textrm{B}}}{2-x_{\textrm{B}}}\left(  \frac{x_{\textrm{B}}}{2}F_{1} + \frac{t}{4M^{2}}F_{2}
\right)\widetilde{\mathcal{E}}\,,
\end{equation}
\end{center}
where the dominant summand is $F_1\widetilde{\mathcal{H}}$ at HERMES kinematic conditions. The $\mathcal{C}^\text{I}_{\text{LP}}$-function is the only $\mathcal{C}$-function that is dominated by CFF $\widetilde{\mathcal{H}}$, and therefore GPD $\widetilde{H}$. Consequently, the asymmetry amplitudes shown in Table~\ref{tab:coefficients} that provide access to $\mathcal{C}_{\text{LP}}^{\text{I}}$ offer the best possibility to constrain~$\widetilde{H}$.

\TABLE[t]{

\begin{tabular}[c]{|c||c|c|c|c|}
\cline{1-5}Asymmetry & Contributory Fourier- & Power of $\frac{1}{Q}$ & Dominant CFF  & Twist\\ 
           Amplitude & Coefficients  & Suppression & Dependence & Level\\
\cline{1-5}\hline
\cline{1-5}\hline \hline \multirow{2}{*}{$A_{\text{UL}}^{\sin\phi}$} & $s_{1,\textrm{LP}}^{\textrm{I}}$ & 1 & \text{Im}\,$\mathcal{C}_{\textrm{LP}}^{\textrm{I}}$ & 2\\
                                          & $s_{1,\textrm{LP}}^{\text{DVCS}}$ & 2& \text{Im}\,$\mathcal{C}_{\textrm{LP}}^{\text{DVCS}}$ & 3\\
\cline{1-5} \multirow{2}{*}{$A_{\text{UL}}^{\sin(2\phi)}$} & $s_{2,\textrm{LP}}^{\textrm{I}}$ & 2 & \text{Im}\,$\mathcal{C}_{\textrm{LP}}^{\textrm{I}}$  & 3\\
                                 & $s_{2,\textrm{LP}}^{\text{DVCS}}$  & 2 & \text{Im}\,$\mathcal{C}_{\textrm{T},\textrm{LP}}^{\text{DVCS}}$ & 2\\
\cline{1-5}$A_{\text{UL}}^{\sin(3\phi)}$ & $s_{3,\textrm{LP}}^{\textrm{I}}$ & 1 & $\text{Im}\,\mathcal{C}_{\textrm{T},\textrm{LP}}^{\textrm{I}}$  & 2 \\
\cline{1-5}\hline \hline \multirow{2}{*}{$A_{\text{LL}}^{\cos(0\phi)}$} &  $c_{0,\textrm{LP}}^{\textrm{I}}$ & 1 & \text{Re}\,$\mathcal{C}_{\textrm{LP}}^{\textrm{I}}$  & 2\\
                                              & $c_{0,\textrm{LP}}^{\text{DVCS}}$  & 1 & \text{Re}\,$\mathcal{C}_{\textrm{LP}}^{\text{DVCS}}$ & 2 \\
\cline{1-5}\multirow{2}{*}{$A_{\text{LL}}^{\cos\phi}$} & $c_{1,\textrm{LP}}^{\textrm{I}}$ & 1 & \text{Re}\,$\mathcal{C}_{\textrm{LP}}^{\textrm{I}}$ & 2\\
                             & $c_{1,\textrm{LP}}^{\text{DVCS}}$  & 3 & \text{Re}\,$\mathcal{C}_{\textrm{LP}}^{\text{DVCS}}$ & 3\\
\cline{1-5}$A_{\text{LL}}^{\cos(2\phi)}$ & $c_{2,\textrm{LP}}^{\textrm{I}}$ & 2 & \text{Re}\,$\mathcal{C}_{\textrm{LP}}^{\textrm{I}}$ & 3\\
\cline{1-5}
\end{tabular}
\label{tab:coefficients}
\caption{The correspondences between the asymmetry
amplitudes extracted from the data set and the Compton form factor dependent Fourier
coefficients of the differential cross-section. The subscript {\sc t} refers to $\mathcal{C}$-functions that involve gluon transversity~\cite{Belitsky2002} and are further suppressed by $\frac{\alpha_{s}}{\pi}$.
}

}
Examination of Table~\ref{tab:coefficients} reveals three asymmetry amplitudes ($A_{\textrm{LL}}^{\cos(0\phi)}, A_{\textrm{UL}}^{\sin\phi}, A_{\textrm{LL}}^{\cos\phi}$) that have leading-twist contributions from the $\mathcal{C}_{\textrm{LP}}^{\textrm{I}}$-function, via the Fourier coefficients ($c_{\textrm{0,LP}}^{\textrm{I}}, s_{\textrm{1,LP}}^{\textrm{I}}, c_{\textrm{1,LP}}^{\textrm{I}}$) in the numerators of Eqs.~\ref{eqn:AUL_effective} and~\ref{eqn:ALL_effective}. The $A_{\text{LL}}^{\cos(0\phi)}$ amplitude receives an additional twist-2 contribution from the $\mathcal{C}_{\text{LP}}^{\text{DVCS}}$-function, stemming from the $c_{0,\text{LP}}^{\text{DVCS}}$ Fourier coefficient.
The $A_{\textrm{LL}}^{\cos(0\phi)}$ and $A_{\text{LL}}^{\cos\phi}$ amplitudes also receive a contribution from BH coefficients, $c_{0,\text{LP}}^{\text{BH}}$ and $c_{1,\text{LP}}^{\text{BH}}$ respectively. The tangled mix of contributions to these amplitudes increases the difficulty of extracting information related to CFFs and therefore GPDs from it. The $A_{\textrm{UL}}^{\sin\phi}$ and $ A_{\textrm{LL}}^{\cos\phi}$ asymmetry amplitudes each receive contributions at the twist-2 level from the $\mathcal{C}^{\text{I}}_{\text{LP}}$-function, with a twist-3 contribution from the $\mathcal{C}_{\textrm{LP}}^{\textrm{DVCS}}$-function. The dominance of the Interference term over the squared-DVCS term, combined with the additional $\frac{1}{Q}$-suppression arising from the twist-3 nature of the $\mathcal{C}_{\textrm{LP}}^{\textrm{DVCS}}$ contribution to these asymmetry amplitudes, makes $A_{\textrm{UL}}^{\sin\phi}$ and $ A_{\textrm{LL}}^{\cos\phi}$ the simplest of the asymmetry amplitudes considered in this paper from which to extract GPD-related information. More specifically, it is expected that these measurements can be used to constrain the values of the CFF $\widetilde{\mathcal{H}}$. The $A^{\sin\phi}_{\text{UL}}$ amplitude is sensitive to the imaginary part of the CFF $\widetilde{\mathcal{H}}$, whereas the $A_{\text{LL}}^{\cos\phi}$ amplitude is sensitive to the real part of the same CFF.

The higher order Fourier components of the asymmetries receive twist-3 or gluon-transversity contributions, or a combination thereof; the $A_{\textrm{UL}}^{\sin(2\phi)}$ and $A_{\textrm{LL}}^{\cos(2\phi)}$ both have twist-3 contributions from the imaginary and real parts of $\mathcal{C}_{\textrm{LP}}^{\textrm{I}}$ respectively. The $A_{\textrm{UL}}^{\sin(2\phi)}$ amplitude also has a contribution from the gluon-transversity-dependent $\mathcal{C}_{\textrm{T,LP}}^{\textrm{DVCS}}$~\cite{Belitsky2002}, while the $A_{\textrm{UL}}^{\sin(3\phi)}$ amplitude depends on a gluon-transversity function from the Interference term, $\mathcal{C}_{\textrm{T,LP}}^{\textrm{I}}$. Both gluon-transversity functions are suppressed by $\frac{\alpha_{s}}{\pi}$ compared to the quark leading-twist case and are expected to be very small. To date these gluon contributions have not been included in any model predictions for any target state. 

\section{Background Correction and Systematic Uncertainties}
The extracted asymmetry amplitudes are corrected in each kinematic bin for background contributions from semi-inclusive and exclusive neutral meson
production. The method is described in Ref.~\cite{Airapetian2008} and the corrected asymmetry amplitude $A_{\text{corr}}$ is given by

\begin{equation}\label{eqn:backcorr}
A_{\text{corr}} = \frac{A -f_{\text{semi}} A_{\text{semi}} - f_{\text{excl}}
A_{\text{excl}}}  {1 - f_{\text{semi}} - f_{\text{excl}}  }.
\end{equation}
The fractional contributions to the data yield from semi-inclusive neutral mesons $f_{\text{semi}}$ and exclusive pions $f_{\text{excl}}$ are estimated from MC simulations~\cite{Vanderhaeghen1999}. The average semi-inclusive contribution is 3.1$\%$, with the greatest variation being across the projection in $x_{\text{B}}$, from 1.1\,\% to 8.0\,\%. The exclusive neutral pion contribution is less than 0.7$\%$ in every kinematic bin. This estimate of the exclusive fraction is supported by calculations from another model~\cite{Goloskokov:2008ib}, and a data search for corresponding events at HERMES~\cite{Vandenbroucke}.

The background asymmetry amplitude from the semi-inclusive process $A_{\text{semi}}$ is extracted from data by reconstructing neutral pions from the two-photon decay, requiring a fractional energy $\text{E}_{\pi^{0}}/\nu > 0.8$ and an invariant mass\, 0.10\,GeV~$< M_{\gamma\gamma} <$~0.17\,GeV. The asymmetry amplitude from the exclusive background $A_{\text{excl}}$ cannot be extracted from data due to the small yield of exclusive neutral pions.  Thus they are assumed to be zero with an uncertainty of $\pm\frac{2}{\sqrt{12}}$, corresponding to one standard deviation from a uniform distribution in the range $[-1,1]$. Half the effect of the total background correction is assigned as a systematic uncertainty. 
The statistical uncertainties of the background fractions and asymmetry amplitudes appearing in Eq.~\ref{eqn:backcorr} are propagated through to the final statistical uncertainty of the amplitude.

The resulting corrected asymmetry amplitudes presented in this paper contain contributions from the elastic BH/DVCS processes and processes in which the proton is excited to a resonant state.  An MC simulation using a parametrisation of the form factor for the resonance region from Ref.~\cite{Brasse1976} is used to estimate the fractional contribution of this resonance process and the individual cross-sections for single-meson decay channels are calculated with the MAID2000 program~\cite{MAID2000}. The contribution to the exclusive sample averages 13$\%$, with the greatest variation of between $5.6\,\%$ and $33.6\,\%$ being across the $t$-range covered in this analysis. We assign no systematic uncertainty due to this contribution and no correction is applied; this contribution is considered to be part of the signal.

The extracted asymmetries are also subject to systematic uncertainties arising
from the combined effects of detector misalignment, acceptance, smearing and
finite bin width, in addition to the background correction described
previously.  The systematic uncertainty originating from the combined
contribution is estimated from a simulation of the spectrometer using a MC generator based on GPD parametrisation detailed in
Ref.~\cite{Korotkov2002}. The resultant uncertainty is denoted ``Detector/Binning Effects'' and is shown in Table~\ref{tab:overallsyst}.

Finally, we assign a systematic uncertainty due to the year-dependent shift in the exclusive region of the missing-mass distribution. We choose to assign one quarter of the effect that the shift has on the extracted asymmetry amplitudes.  

No systematic uncertainty is assigned due to luminosity differences because the luminosity does not depend on the target polarisation and beam polarisation dependent weights are assigned to each event in the extraction.  Possible uncertainties arising from extra QED vertices are neglected as the effects have been estimated to be negligible for HERMES~\cite{afanasev}.  

The magnitudes of all the contributions discussed in this section are given in Table~\ref{tab:overallsyst}.

\TABLE[t]{

\begin{tabular}[c]{|c||r||c|c|c|}
\cline{1-5} & \multicolumn{1}{c||}{} &  Missing &  & Detector/  \\ 
Amplitude & \multicolumn{1}{c||}{A\,$\pm\,\delta_{\text{stat.}}\,\pm\,\delta_{\text{syst.}}$} &  Mass & Background & Binning \\ 
 & \multicolumn{1}{c||}{} &  Shift &  & Effects  \\ \cline{1-5}
\hline \hline
$A_{\text{UL}}^{\sin\phi}$ & -0.073\,$\pm$\,0.032\,$\pm$\,0.007 & 0.002 & 0.006 & 0.002 \\ \cline{1-5}
$A_{\text{UL}}^{\sin(2\phi)}$ & -0.107\,$\pm$\,0.032\,$\pm$\,0.008 & 0.000 & 0.002 & 0.007 \\ \cline{1-5}
$A_{\text{UL}}^{\sin(3\phi)}$ & 0.015\,$\pm$\,0.032\,$\pm$\,0.009 & 0.000 & 0.001 & 0.009\\ \cline{1-5}
\hline \hline
$A_{\text{LL}}^{\cos(0\phi)}$ & 0.122\,$\pm$\,0.044\,$\pm$\,0.004 & 0.001 & 0.003 & 0.003\\ \cline{1-5}
$A_{\text{LL}}^{\cos\phi}$ & -0.047\,$\pm$\,0.062\,$\pm$\,0.029 & 0.002 & 0.003 & 0.028\\ \cline{1-5}
$A_{\text{LL}}^{\cos(2\phi)}$ & 0.067\,$\pm$\,0.062\,$\pm$\,0.007 & 0.004 & 0.003 & 0.004\\ \cline{1-5}
\end{tabular}
\caption{Results for the asymmetry amplitudes extracted over the kinematic range covered by HERMES, with their statistical and systematic uncertainties. For the latter, the various systematic uncertainty contributions are also presented.  Not included are scale uncertainties of 4.2\% (5.3\%) arising from the target (beam and target) polarisation measurements.}
\label{tab:overallsyst}
}

\section{Results and Model Comparison}
Figures~\ref{fig:H_ul_c0s1s2} and~\ref{fig:H_ll_c0c1c2} respectively show the asymmetry amplitudes  $A_{\text{UL}}^{\sin(n\phi)}$ and $A_{\text{LL}}^{\cos(n\phi)}$ integrated over the HERMES acceptance as well as projected across the kinematic variables $-t$, $x_\text{B}$ and $Q^2$. All values are summarised in Table~\ref{tab:Aul_ll}.  It should be noted that, within the HERMES acceptance, $\langle x_{\text{B}} \rangle$ and $\langle Q^2 \rangle$ are highly correlated (see Table~\ref{tab:Aul_ll}) and it is not possible to disentangle the dependences of the asymmetries on these quantities.  The measurements are also subject to scale uncertainties from the measurements of the beam and/or target polarisations. These scale uncertainties are given in the captions to the figures.  The fractional contribution to the data set from resonant state production ($i.e.$ ``Reso. frac.'') is estimated from MC simulations and shown in the bottom row of Figs.~\ref{fig:H_ul_c0s1s2} and ~\ref{fig:H_ll_c0c1c2}. It is not known how this contribution may affect the values of the extracted asymmetries and since here it is not experimentally separable from the non-resonant data, it is treated as a part of the signal. 

\FIGURE{
\includegraphics[width=\textwidth]{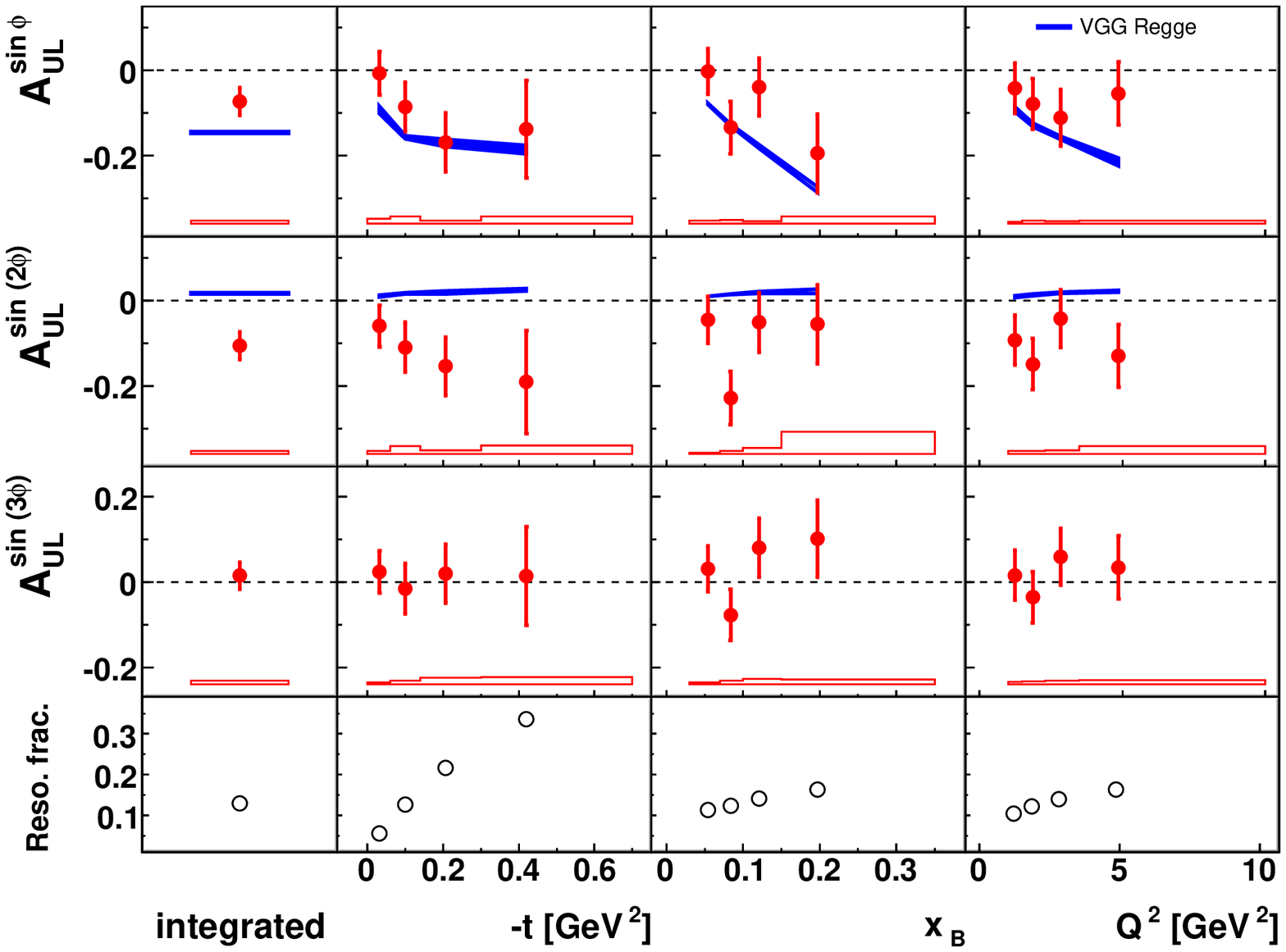}
\caption{Amplitudes of the target-spin asymmetry $\mathcal{A}_{\text{UL}}$ sensitive to a combination of the Interference and squared-DVCS terms, for positrons incident on longitudinally polarised protons, as projections in $-t$, $x_{\text{B}}$, and $Q^{2}$. The leftmost column shows the asymmetry values when the extraction is performed in a single bin across the entire kinematic range of the data set. The error bars (open red bands) show the statistical (systematic) uncertainties and the solid blue bands represent the predictions from the ``VGG Regge'' GPD model described in Refs.~\cite{Vanderhaeghen1999,Vanderhaegen2001}. There is an additional 4.2\% scale uncertainty due to the precision of the measurement of the target polarisation. The fractional contributions from resonance production estimated from an MC model are presented in the bottom panel.}
\label{fig:H_ul_c0s1s2}
}

\FIGURE{
\includegraphics[width=\textwidth]{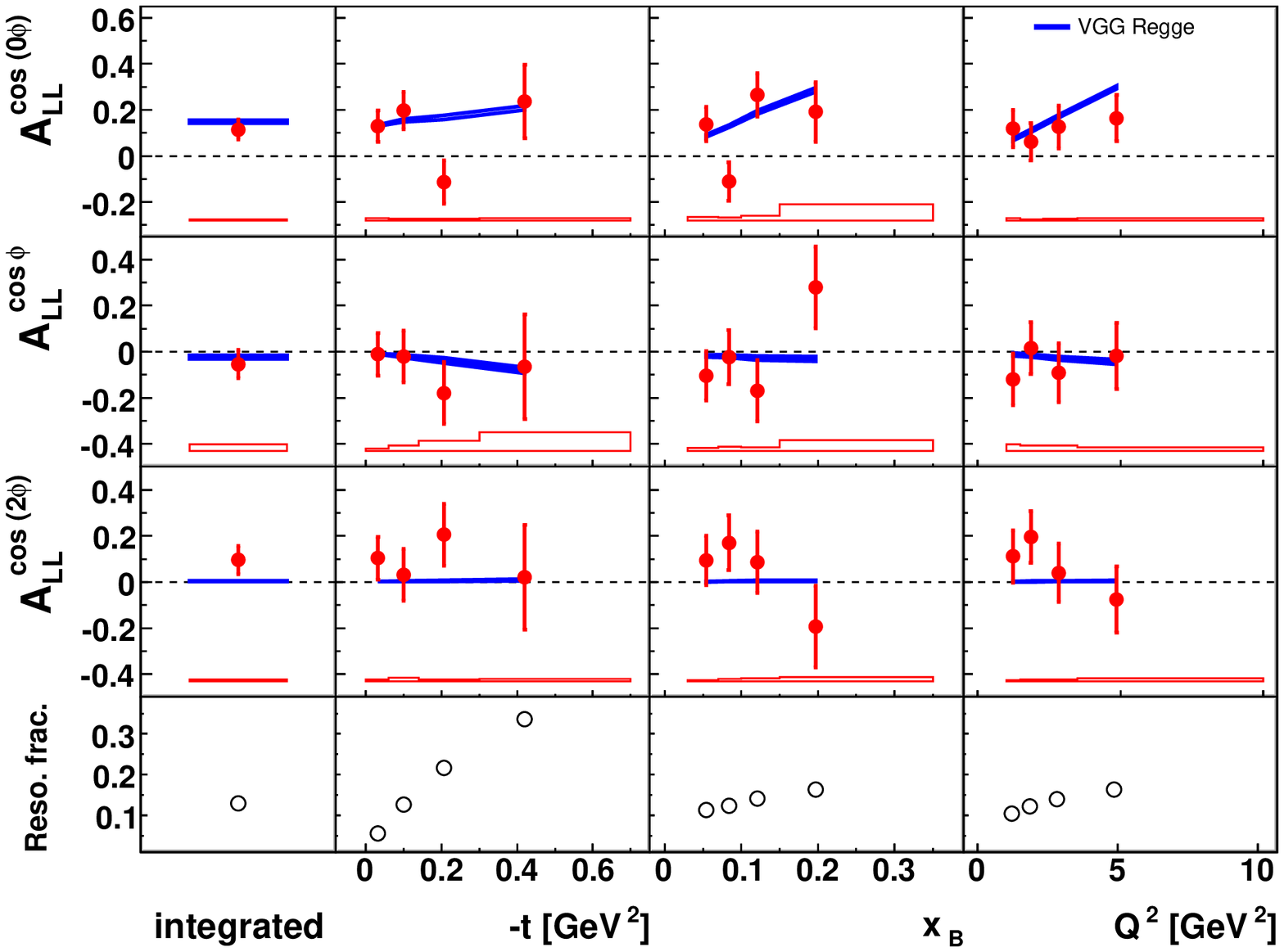}
\caption{Amplitudes of the double-spin asymmetry $\mathcal{A}_{\text{LL}}$ sensitive to the Interference, squared-DVCS and squared-BH terms in the scattering amplitude, for polarised positrons incident on longitudinally polarised protons, as projections in $-t$, $x_{\text{B}}$, and $Q^{2}$.  The leftmost column shows the asymmetry values when the extraction is performed in a single bin across the entire kinematic range of the data set. The error bars (open red bands) show the statistical (systematic) uncertainties and the solid blue bands represent the theoretical predictions from the ``VGG Regge'' GPD model described in Ref.~\cite{Vanderhaeghen1999,Vanderhaegen2001}. There is an additional 5.3\% scale uncertainty due to the precision of the measurement of the beam and target polarisations. The fractional contributions from resonance production estimated from an MC model are presented in the bottom panel.}
\label{fig:H_ll_c0c1c2}
}

All amplitudes presented correspond to Fourier coefficients described in Ref.~\cite{Belitsky2002} relating to the twist-2 and twist-3 CFFs shown in Table~\ref{tab:coefficients} with the caveat that this relationship may be complicated by various $c_n$ terms in the denominators of Eqs.~\ref{eqn:AUL_effective} and~\ref{eqn:ALL_effective}. 

The first harmonic of the $\mathcal{A}_{\text{UL}}$, when the extraction is performed in a single bin from all kinematics, exhibits the value $A_{\text{UL}}^{\sin\phi}=-0.073\pm0.032\,(\text{stat.})\pm 0.007\,(\text{syst.})$. The kinematic projections provide no evidence of strong dependences on $-t$, $x_\text{B}$, or $Q^2$. This asymmetry amplitude receives a mixture of twist-2 and twist-3 contributions, as shown in Table~\ref{tab:coefficients}. The primary contributor is $\mathcal{C}^{\text{I}}_{\text{LP}}$, which is twist-2 and is expected to dominate the twist-3 contribution from $\mathcal{C}^{\text{DVCS}}_{\text{LP}}$.

The $A_\text{UL}^{\sin(2\phi)}$ amplitude has the unexpectedly large value $A_\text{UL}^{\sin(2\phi)}=-0.106\pm0.032\pm0.008$ when extracted from the integrated kinematic range of the data set. The projections across $-t$, $x_{\text{B}}$ and $Q^2$ in Fig.~\ref{fig:H_ul_c0s1s2} show no obvious features. This asymmetry amplitude is expected to receive a mixture of quark twist-3 and gluon twist-2 contributions, and as such could have been expected to be small in the HERMES kinematic range.

The CLAS collaboration also published extractions~~\cite{Chen2006} of $A^{\sin\phi}_{\text{UL}}$ and $A^{\sin(2\phi)}_{\text{UL}}$ although without projections in $-t$, $x_{\text{B}}$ and $Q^2$ across the kinematic region covered by CLAS. In Table~\ref{tab:hermesclas}, the signs of the CLAS results are suitable for comparison with HERMES results. The sign of the $A_{\text{UL}}^{\sin\phi}$ ($A_{\text{UL}}^{\sin(2\phi)}$) amplitude has been inverted once (twice) to account for the different beam charge used and to bring the angular definitions used by CLAS into concordance with the Trento convention used by HERMES~\cite{Bacchetta2004}. The single bin extraction value evaluated by CLAS has approximately three times larger magnitude than that evaluated by HERMES, whereas the $A^{\sin(2\phi)}_{\text{UL}}$ asymmetry amplitude at CLAS is consistent with zero in contrast to the non-zero value extracted by HERMES. It should be noted that CLAS measurements are taken with a different beam charge and at larger average values of $x_{\text{B}}$ and $t$ than HERMES, resulting in a different sensitivity to CFFs $\mathcal{H}$ and $\widetilde{\mathcal{H}}$ as can be deduced from Eq.~\ref{eqn:twist-2_C_LP}.

The $A_{\text{UL}}^{\sin(3\phi)}$ and consistency test $A_{\text{UL}}^{\cos(0\phi)}$ amplitudes were found to be consistent with zero over the kinematic range of the HERMES experiment. The former receives contributions from the leading-twist $\mathcal{C}_{\text{T,LP}}^{\text{I}}$-function. 

The results from the extraction of the double-spin asymmetry are presented in Fig.~\ref{fig:H_ll_c0c1c2}. The twist-2 amplitude $A_{\text{LL}}^{\cos(0\phi)}$ is found to be $0.115\pm0.044\pm0.004$ when the extraction is performed in a single bin across the entire kinematic range of the data set. No dependences are observed in the data across projections in $-t, x_\text{B}$, or $Q^2$. This asymmetry amplitude receives contributions from the squared-DVCS and Interference terms in the scattering amplitude, as shown in Table~\ref{tab:coefficients}. However, it also receives a dominating contribution from the $c^{\text{BH}}_{0,\text{LP}}$ Fourier coefficient as shown in Eq.~\ref{eqn:ALL_effective}. It is therefore expected to be positive and non-zero, as confirmed by the data.

The first harmonic of the double-spin asymmetry $A_{\text{LL}}^{\cos\phi}=-0.054\pm0.062\pm0.029$  is compatible with zero when the extraction is performed in a single bin across the entire kinematic range of the data set. It exhibits no dependences on $-t, x_\text{B}$, or $Q^2$. It is expected to receive contributions from twist-2 and twist-3 terms, as shown in Table~\ref{tab:coefficients}, and therefore may be non-zero at HERMES kinematics.  Its value is dominated by the $c^{\text{BH}}_{1,\text{LP}}$ term arising from the squared-BH term in the expansion of the scattering amplitude.

The $A_{\text{LL}}^{\cos(2\phi)}$ amplitude is also compatible with zero. The kinematic projections show no dependence on $-t, x_\text{B}$, or $Q^2$. This asymmetry amplitude receives only twist-3 contributions, and so is expected to be small at HERMES kinematics. Unlike the single-spin asymmetry, there are no previous experimental measurements with which to compare the extracted double-spin asymmetry.

All asymmetry amplitudes in Figs.~\ref{fig:H_ul_c0s1s2} and \ref{fig:H_ll_c0c1c2} are presented in comparison with calculations~\cite{Vanderhaegen2001}, labelled ``VGG Regge'', from the GPD model described in Ref.~\cite{Vanderhaeghen1999}. Predictions from this model have been compared previously with HERMES asymmetries with respect to beam helicity and beam charge~\cite{Airapetian2009} and for a transversely polarised target~\cite{Airapetian2008} with limited success for certain choices of parameters. It remains the only predictive model available for comparison with data taken on a polarised target. The model is an implementation of the double distribution concept~\cite{D.Mueller1994,Radyushkin1996} where the kernel of the double distribution contains a profile function that determines the dependence on $\xi$, controlled by a parameter $b$~\cite{Musatov2000}. The $b$ profile parameter can be set in the range $b\in[1,\infty)$, where the GPD is independent of $\xi$ in the limit $b\rightarrow\infty$. In this model, the $\xi$-dependence and the $x$-dependence are entangled, but the $t$-dependence is factorised. Here a Regge-inspired hypothesis for the $t$-dependence is used.

The model is used to calculate the differential cross-sections for the electroproduction of real photons for each beam and target polarisation state. These differential cross-sections are used to construct asymmetries with which to compare the data. The cross-sections and therefore the asymmetries contain no provision for the production of resonant states of the target proton. The width of the theory bands in Figs.~\ref{fig:H_ul_c0s1s2} and \ref{fig:H_ll_c0c1c2} originates from the variation of the unknown $b$ profile parameters. These free parameters are independently controllable for valence and sea quarks, and can be used as a fit parameter for the extraction of GPDs from hard leptoproduction data. The other parameters of the model are chosen from those that best describe data that were previously published by HERMES~\cite{Airapetian2009}.

Figure~\ref{fig:H_ul_c0s1s2} shows that the reported amplitude $A_{\text{UL}}^{\sin\phi}$ of the $\mathcal{A}_{\text{UL}}$ single-spin asymmetry is well described by the model, which predicts the kinematic trend and the magnitude of this amplitude with reasonable accuracy. However, the relatively large amplitude $A_{\text{UL}}^{\sin(2\phi)}$ is not described by the model, which predicts that the amplitude should be small and of opposite sign. Although Ref.~\cite{Chen2006} alludes to a large exclusive pion background in the CLAS kinematic region, which could cause a large $\sin(2\phi)$ amplitude, as previously stated a data search at HERMES~\cite{Vandenbroucke} showed no such contamination. The model~\cite{Vanderhaeghen1999} itself suggests that any background due to exclusive pions should be small and this is supported by a different set of calculations from a different model~\cite{Goloskokov:2008ib}. This disagreement between the data and the model for this amplitude is surprising and could be interpreted as an unexpectedly large contribution from the gluon twist-2 amplitudes, which are not included in the GPD model shown.

Figure~\ref{fig:H_ll_c0c1c2} shows that the predictions made by the model regarding the magnitude and trends of the amplitudes of the $\mathcal{A}_{\text{LL}}$ asymmetry mostly agree with the data.  It describes the positive, slightly increasing trend for the $A_{\text{LL}}^{\cos(0\phi)}$ amplitude observed in the data across all three variables. The prediction of the model for values of the $A_{\text{LL}}^{\cos\phi}$ is compatible with the data within the uncertainties of the extraction. The model predicts a small value for $A_{\text{LL}}^{\cos(2\phi)}$ across all kinematic projections which is compatible with zero within experimental uncertainty. 

The model is successful in predicting four of the five experimental asymmetries presented in this paper. However, the same model failed to predict previously published HERMES results~\cite{Airapetian2009} for the beam helicity asymmetry. A further caveat is that the model does not include any effect from the resonance contribution to the data set.

\TABLE[t]{
\centering
\tiny
\begin{tabular}{|c||c|c|c||r|r|r|}
\cline{1-7} & & & & \multicolumn{1}{c|}{} & \multicolumn{1}{c|}{}& \multicolumn{1}{c|}{}\\
Kinematic & $\langle -t \rangle$ & $\langle x_{\text{B}} \rangle$ & $\langle Q^{2} \rangle$ & \multicolumn{1}{c|}{$A^{\sin{\phi}}_{\text{UL}}$} & \multicolumn{1}{c|}{$A^{\sin(2\phi)}_{\text{UL}}$} & \multicolumn{1}{c|}{$A^{\sin(3\phi)}_{\text{UL}}$}\\
Bin & [GeV$^{2}$]&- & [GeV$^{2}$]& \multicolumn{1}{c|}{$\pm\,{\delta}_{\text{stat.}}\,\pm\,{\delta}_{\text{syst.}}$} & \multicolumn{1}{c|}{$\pm\,{\delta}_{\text{stat.}}\,\pm\,{\delta}_{\text{syst.}}$} & \multicolumn{1}{c|}{$\pm\,{\delta}_{\text{stat.}}\,\pm\,{\delta}_{\text{syst.}}$} \\
& & & & \multicolumn{1}{c|}{} & \multicolumn{1}{c|}{} & \multicolumn{1}{c|}{} \\ \cline{1-7}
\hline \hline integrated & $0.115$ & $0.096$ & $2.459$ & $-0.073\pm0.032\pm0.007$ & $-0.106\pm0.032\pm0.008$ & $0.015\pm0.032\pm0.009$\\
\hline $0.00 \leq -t\leq0.06$ & $0.031$ & $0.079$ & $1.982$ & $-0.008\pm0.051\pm0.012$ & $-0.060\pm0.050\pm0.007$ & $0.024\pm0.049\pm0.005$\\
$0.06<-t\leq0.14$ & $0.094$ & $0.103$ & $2.531$ & $-0.085\pm0.057\pm0.017$ & $-0.110\pm0.059\pm0.018$ & $-0.016\pm0.059\pm0.009$\\
$0.14<-t\leq0.30$ & $0.201$ & $0.110$ & $2.883$ & $-0.169\pm0.070\pm0.007$ & $-0.154\pm0.069\pm0.008$ & $0.020\pm0.069\pm0.016$\\
$0.30<-t\leq0.70$ & $0.408$ & $0.123$ & $3.587$ & $-0.138\pm0.109\pm0.017$ & $-0.191\pm0.116\pm0.021$ & $0.014\pm0.115\pm0.017$\\
\hline
$0.03<{x_{\text{B}}}\leq0.07$ & $0.096$ & $0.054$ & $1.437$ & $-0.003\pm0.053\pm0.008$ & $-0.045\pm0.053\pm0.004$ & $0.031\pm0.053\pm0.004$\\
$0.07<{x_{\text{B}}}\leq0.10$ & $0.099$ & $0.084$ & $2.115$ & $-0.134\pm0.064\pm0.008$ & $-0.228\pm0.064\pm0.007$ & $-0.077\pm0.060\pm0.009$\\
$0.10<{x_{\text{B}}}\leq0.15$ & $0.123$ & $0.121$ & $3.108$ & $-0.039\pm0.070\pm0.007$ & $-0.051\pm0.069\pm0.015$ & $0.080\pm0.069\pm0.013$\\
$0.15<{x_{\text{B}}}\leq0.35$ & $0.188$ & $0.198$ & $4.934$ & $-0.195\pm0.093\pm0.018$ & $-0.056\pm0.089\pm0.052$ & $0.101\pm0.090\pm0.012$\\
\hline
$1.0<{Q^{2}}\leq1.5$ & $0.085$ & $0.056$ & $1.236$ & $-0.043\pm0.059\pm0.004$ & $-0.093\pm0.059\pm0.007$ & $0.016\pm0.058\pm0.006$\\
$1.5<{Q^{2}}\leq2.3$ & $0.098$ & $0.079$ & $1.862$ & $-0.079\pm0.060\pm0.007$ & $-0.149\pm0.061\pm0.007$ & $-0.036\pm0.060\pm0.007$\\
$2.3<{Q^{2}}\leq3.5$ & $0.123$ & $0.108$ & $2.829$ & $-0.111\pm0.068\pm0.007$ & $-0.042\pm0.067\pm0.009$ & $0.059\pm0.066\pm0.009$\\
$3.5<{Q^{2}}\leq10$ & $0.178$ & $0.170$ & $4.865$ & $-0.054\pm0.071\pm0.008$ & $-0.130\pm0.074\pm0.019$ & $0.034\pm0.074\pm0.010$\\ \cline{1-7}
\hline\hline
\cline{1-7} & & & & \multicolumn{1}{c|}{} & \multicolumn{1}{c|}{} & \multicolumn{1}{c|}{}\\
Kinematic & $\langle -t \rangle$ & $\langle x_{\text{B}} \rangle$ & $\langle Q^{2} \rangle$ & \multicolumn{1}{c|}{$A^{\cos(0\phi)}_{\text{LL}}$} & \multicolumn{1}{c|}{$A^{\cos{\phi}}_{\text{LL}}$} & \multicolumn{1}{c|}{$A^{\cos(2\phi)}_{\text{LL}}$} \\
Bin & [GeV$^{2}$] & - & [GeV$^{2}$] &  \multicolumn{1}{c|}{$\pm\,{\delta}_{\text{stat.}}\,\pm\,{\delta}_{\text{syst.}}$} & \multicolumn{1}{c|}{$\pm\,{\delta}_{\text{stat.}}\,\pm\,{\delta}_{\text{syst.}}$} & \multicolumn{1}{c|}{$\pm\,{\delta}_{\text{stat.}}\,\pm\,{\delta}_{\text{syst.}}$} \\
& & & & \multicolumn{1}{c|}{} & \multicolumn{1}{c|}{} & \multicolumn{1}{c|}{}\\ \cline{1-7}
\hline \hline integrated & $0.115$ & $0.096$ & $2.459$ & $0.115\pm0.044\pm0.004$ & $-0.054\pm0.062\pm0.029$ & $0.095\pm0.062\pm0.007$ \\
\hline $0.00 \leq -t\leq0.06$ & $0.031$ & $0.079$ & $1.982$ & $0.129\pm0.068\pm0.010$ & $-0.012\pm0.094\pm0.010$ & $0.104\pm0.097\pm0.006$ \\
$0.06<-t\leq0.14$ & $0.094$ & $0.103$ & $2.531$ & $0.197\pm0.080\pm0.007$ & $-0.021\pm0.112\pm0.022$ & $0.031\pm0.114\pm0.014$ \\
$0.14<-t\leq0.30$ & $0.201$ & $0.110$ & $2.883$ & $-0.113\pm0.095\pm0.008$ & $-0.179\pm0.137\pm0.044$ & $0.206\pm0.136\pm0.007$ \\
$0.30<-t\leq0.70$ & $0.408$ & $0.123$ & $3.587$ & $0.237\pm0.162\pm0.009$ & $-0.065\pm0.235\pm0.079$ & $\,0.020\pm0.211\pm0.009$ \\
\hline
$0.03<{x_{\text{B}}}\leq0.07$ & $0.096$ & $0.054$ & $1.437$ & $0.137\pm0.076\pm0.014$ & $-0.108\pm0.108\pm0.013$ & $0.094\pm0.102\pm0.003$ \\
$0.07<{x_{\text{B}}}\leq0.10$ & $0.099$ & $0.084$ & $2.115$ & $-0.111\pm0.083\pm0.012$ & $-0.023\pm0.123\pm0.018$ & $0.171\pm0.117\pm0.008$ \\
$0.10<{x_{\text{B}}}\leq0.15$ & $0.123$ & $0.121$ & $3.108$ & $0.265\pm0.095\pm0.020$ & $-0.169\pm0.135\pm0.016$ & $0.087\pm0.132\pm0.012$ \\
$0.15<{x_{\text{B}}}\leq0.35$ & $0.188$ & $0.198$ & $4.934$ & $0.192\pm0.125\pm0.070$ & $0.279\pm0.178\pm0.047$ & $-0.193\pm0.176\pm0.017$ \\
\hline
$1.0<{Q^{2}}\leq1.5$ & $0.085$ & $0.056$ & $1.236$ & $0.118\pm0.080\pm0.009$ & $-0.120\pm0.114\pm0.029$ & $0.111\pm0.115\pm0.004$ \\
$1.5<{Q^{2}}\leq2.3$ & $0.098$ & $0.079$ & $1.862$ & $0.061\pm0.082\pm0.005$ & $0.015\pm0.120\pm0.024$ & $0.196\pm0.113\pm0.006$ \\
$2.3<{Q^{2}}\leq3.5$ & $0.123$ & $0.108$ & $2.829$ & $0.126\pm0.091\pm0.008$ & $-0.092\pm0.130\pm0.023$ & $0.040\pm0.130\pm0.005$ \\
$3.5<{Q^{2}}\leq10$ & $0.178$ & $0.170$ & $4.865$ & $0.164\pm0.102\pm0.010$ & $-0.019\pm0.149\pm0.016$ & $-0.076\pm0.149\pm0.013$ \\ \cline{1-7}
\end{tabular}
\caption{Results of the $A_{\text{UL}}^{\sin\phi}$, $A_{\text{UL}}^{\sin(2\phi)}$ and $A_{\text{UL}}^{\sin(3\phi)}$ amplitudes of the single-spin asymmetry (top) and $A_{\text{LL}}^{\cos(0\phi)}$, $A_{\text{LL}}^{\cos\phi}$, and $A_{\text{LL}}^{\cos(2\phi)}$ amplitudes of the double-spin asymmetry (bottom) with statistical and systematic uncertainties and average kinematics for polarised hydrogen data.  These are shown integrated over the HERMES acceptance and for each $-t$, $x_{\text{B}}$, and $Q^{2}$ bin. The correlation between $\langle x_{\text{B}}\rangle$ and $\langle Q^2\rangle$ can be observed in the average kinematic values for each bin. There is an additional scale uncertainty of 4.2\%\,(5.3\%) affecting the $A_{\text{UL}}^{\sin(n\phi)}$ ($A_{\text{LL}}^{\cos(n\phi)}$) amplitudes due to the uncertainty in the measurement of the target (beam and target) polarisations.}
\label{tab:Aul_ll}
}
\normalsize

\TABLE[t]{
\begin{tabular}{|c||c|c|c||c|c|}
\cline{1-6}
\multirow{2}{*}{Experiment} & $\langle -t \rangle$ & $\langle x_{\text{B}} \rangle$ & $\langle Q^{2} \rangle$ & \multirow{2}{*}{$A^{\sin{\phi}}_{\text{UL}}\pm\,{\delta}_{\text{stat.}}\,\pm\,{\delta}_{\text{syst.}}$ } & \multirow{2}{*}{$A^{\sin(2\phi)}_{\text{UL}}\pm\,{\delta}_{\text{stat.}}\,\pm\,{\delta}_{\text{syst.}}$} \\
 & [GeV$^{2}$]&- & [GeV$^{2}$]&  &  \\\cline{1-6}
\hline \hline HERMES & $0.12$ & $0.10$ & $2.46$ & $-0.073\pm0.032\pm0.007$ & $-0.106\pm0.032\pm0.008$ \\\cline{1-6}
CLAS & $0.31$ & $0.28$ & $1.82$ & $-0.252\pm0.042\pm0.020$ & $-0.022\pm0.045\pm0.021$ \\\cline{1-6}
\end{tabular}
\caption{The single-spin asymmetry defined in Eq.~\ref{eqn:AUL_defn}, extracted at HERMES and CLAS with the average kinematic values for the respective datasets. The sign of the CLAS $\sin\phi$ result has been inverted to become consistent with the different angular definitions used by HERMES and the different beam charge used in the two measurements, as described in the text.}
\label{tab:hermesclas}
}

\section{Summary}
Data on the hard exclusive electroproduction of real photons from the 1996 and 1997 years of operation of the HERMES experiment are analysed. Two asymmetries in the azimuthal distribution of leptoproduced photons from a longitudinally polarised proton target are presented.  The $\sin\phi$ modulation $A_{\text{UL}}^{\sin\phi}$ of the single-spin asymmetry $\mathcal{A}_{\text{UL}}(\phi)$ is found to be small and negative, with a weak kinematic dependence.  The asymmetries are compared to calculations from the only available GPD model~\cite{Vanderhaeghen1999,Vanderhaegen2001} for these processes on a longitudinally polarised target.  The asymmetry amplitude is broadly compatible with the theoretical predictions from the model. The $A_{\text{UL}}^{\sin(2\phi)}$ amplitude is found to be of opposite sign and larger than expected.  The double-spin asymmetry $\mathcal{A}_{\text{LL}}(\phi)$ is extracted for the first time. It is found to be consistent with the small asymmetry values predicted by the GPD model. Both of the asymmetries have dominant contributions from the real or imaginary parts of the CFF $\widetilde{\mathcal{H}}$, and thus provide potentially the best access to the GPD $\widetilde{H}$ at HERMES kinematics.

\section{Acknowledgements}
We gratefully acknowledge the \desy\ management for its support and the staff
at \desy\ and the collaborating institutions for their significant effort.
This work was supported by the FWO-Flanders and IWT, Belgium;
the Natural Sciences and Engineering Research Council of Canada;
the National Natural Science Foundation of China;
the Alexander von Humboldt Stiftung;
the German Bundesministerium f\"ur Bildung und Forschung (BMBF);
the Deutsche Forschungsgemeinschaft (DFG);
the Italian Istituto Nazionale di Fisica Nucleare (INFN);
the MEXT, JSPS, and G-COE of Japan;
the Dutch Foundation for Fundamenteel Onderzoek der Materie (FOM);
the U.K.~Engineering and Physical Sciences Research Council, 
the Science and Technology Facilities Council,
and the Scottish Universities Physics Alliance;
the U.S.~Department of Energy (DOE) and the National Science Foundation (NSF);
the Russian Academy of Science and the Russian Federal Agency for 
Science and Innovations;
and the Ministry of Economy and the Ministry of Education and Science of 
Armenia.

\bibliographystyle{JHEP.bst}

\begin{thebibliography}{99}

\bibitem{D.Mueller1994}
D.~M\"uller, D.~Robaschik, B.~Geyer, F.-M. Dittes, and J.~Hoercaron, {\it {Wave
  functions, evolution equations and evolution kernels from light ray operators
  of QCD}},  {\em Fortschritte der Physik/Progress of Physics} {\bf 42} (1994)
  101, [\href{http://xxx.lanl.gov/abs/hep-ph/9812448}{{\tt hep-ph/9812448}}].

\bibitem{Ji1997}
X.~Ji, {\it {Gauge-invariant decomposition of nucleon spin}},  {\em Phys. Rev.
  Lett.} {\bf 78} (1997) 610,
  [\href{http://xxx.lanl.gov/abs/hep-ph/9603249}{{\tt hep-ph/9603249}}].

\bibitem{Radyushkin1996}
A.~V. Radyushkin, {\it {Scaling limit of deeply virtual compton scattering}},
  {\em Phys. Lett. B} {\bf 380} (1996) 417,
  [\href{http://xxx.lanl.gov/abs/hep-ph/9604317}{{\tt hep-ph/9604317}}].

\bibitem{PhysRevD.62.071503}
M.~Burkardt, {\it {Impact parameter dependent parton distributions and
  off-forward parton distributions for $\zeta{}\rightarrow{}0$}},  {\em Phys.
  Rev. D} {\bf 62} (2000) 071503,
  [\href{http://xxx.lanl.gov/abs/hep-ph/0005108}{{\tt hep-ph/0005108}}].

\bibitem{Diehl2003}
M.~Diehl, {\it {Generalized parton distributions}},  {\em Phys. Rept.} {\bf
  388} (2003) 41, [\href{http://xxx.lanl.gov/abs/hep-ph/0307382}{{\tt
  hep-ph/0307382}}].

\bibitem{Bluemlein1999}
J.~Bl\"umlein, B.~Geyer, and D.~Robaschik, {\it {The virtual Compton amplitude
  in the generalized Bjorken region: Twist-2 contributions}},  {\em Nuclear
  Physics B} {\bf 560} (1999) 283,
  [\href{http://xxx.lanl.gov/abs/hep-ph/9903520}{{\tt hep-ph/9903520}}].

\bibitem{Belitsky2000}
A.~V. Belitsky and D.~M\"uller, {\it {Off-forward gluonometry}},  {\em Phys.
  Lett. B} {\bf 486} (2000) 369,
  [\href{http://xxx.lanl.gov/abs/hep-ph/0005028}{{\tt hep-ph/0005028}}].

\bibitem{Vanderhaeghen1999}
M.~Vanderhaeghen, P.~A.~M. Guichon, and M.~Guidal, {\it {Deeply virtual
  electroproduction of photons and mesons on the nucleon: Leading order
  amplitudes and power corrections}},  {\em Phys. Rev. D} {\bf 60} (1999)
  094017, [\href{http://xxx.lanl.gov/abs/hep-ph/9910426}{{\tt
  hep-ph/9910426}}].

\bibitem{Bacchetta2004}
A.~Bacchetta, U.~D\char39{}Alesio, M.~Diehl, and C.~A. Miller, {\it
  {Single-spin asymmetries: The Trento conventions}},  {\em Phys. Rev. D} {\bf
  70} (2004) 117504, [\href{http://xxx.lanl.gov/abs/hep-ph/0410050}{{\tt
  hep-ph/0410050}}].

\bibitem{Korotkov2002}
V.~A. Korotkov and W.~D. Nowak, {\it {Future measurements of deeply virtual
  Compton scattering at HERMES}},  {\em Eur. Phys. J. C} {\bf 23} (2002) 455,
  [\href{http://xxx.lanl.gov/abs/hep-ph/0108077}{{\tt hep-ph/0108077}}].

\bibitem{Belitsky2002}
A.~V. Belitsky, D.~M\"uller, and A.~Kirchner, {\it {Theory of deeply virtual
  Compton scattering on the nucleon}},  {\em Nuclear Physics B} {\bf 629}
  (2002) 323, [\href{http://xxx.lanl.gov/abs/hep-ph/0112108}{{\tt
  hep-ph/0112108}}].

\bibitem{Ackerstaff1998a}
{\bf HERMES} Collaboration, K.~Ackerstaff {\em et~al.}, {\it {The HERMES
  spectrometer}},  {\em Nuclear Instruments and Methods A} {\bf 417} (1998)
  230, [\href{http://xxx.lanl.gov/abs/hep-ex/9806008}{{\tt hep-ex/9806008}}].

\bibitem{HERMESCollaboration2005}
{\bf HERMES} Collaboration, A.~Airapetian {\em et~al.}, {\it {The HERMES
  Polarized Hydrogen and Deuterium Gas Target in the HERA Electron Storage
  Ring}},  {\em Nuclear Instruments and Methods A} {\bf 540} (2005) 68,
  [\href{http://xxx.lanl.gov/abs/physics/0408137}{{\tt physics/0408137}}].

\bibitem{Benisch2001}
T.~Benisch {\em et~al.}, {\it {The luminosity monitor of the HERMES experiment
  at DESY}},  {\em Nuclear Instruments and Methods} {\bf 471} (2001) 314.

\bibitem{Airapetian2007b}
{\bf HERMES} Collaboration, A.~Airapetian {\em et~al.}, {\it {The Beam--Charge
  Azimuthal Asymmetry and Deeply Virtual Compton Scattering}},  {\em Phys. Rev.
  D} {\bf 75} (2007) 011103,
  [\href{http://xxx.lanl.gov/abs/hep-ex/0605108}{{\tt hep-ex/0605108}}].

\bibitem{Kopytin}
M.~Kopytin, {\em {Longitudinal target-spin azimuthal asymmetry in deeply-
  virtual Compton scattering}}.
\newblock PhD thesis, Humboldt University Berlin, 2007.

\bibitem{Mahon2010}
D.~Mahon, {\em {Deeply Virtual Compton Scattering Off Longitudinally Polarised
  Protons at HERMES}}.
\newblock PhD thesis, University of Glasgow, submitted March 2010.

\bibitem{Ellinghaus2004}
F.~Ellinghaus, {\em {Beam Charge and Beam Spin Azimuthal Asymmetries in Deeply
  Virtual Compton Scattering}}.
\newblock PhD thesis, Humboldt University Berlin, 2004.

\bibitem{Airapetian2008}
{\bf HERMES} Collaboration, A.~Airapetian {\em et~al.}, {\it {Measurement of
  Azimuthal Asymmetries With Respect To Both Beam Charge and Transverse Target
  Polarization in Exclusive Electroproduction of Real Photons}},  {\em JHEP}
  {\bf 06} (2008) 066, [\href{http://xxx.lanl.gov/abs/0802.2499}{{\tt
  arXiv:0802.2499}}].

\bibitem{Airapetian2009}
{\bf HERMES} Collaboration, A.~Airapetian {\em et~al.}, {\it {Separation of
  contributions from deeply virtual Compton scattering and its interference
  with the Bethe--Heitler process in measurements on a hydrogen target}},  {\em
  JHEP} {\bf 11} (2009) 083, [\href{http://xxx.lanl.gov/abs/0909.3587}{{\tt
  arXiv:0909.3587}}].

\bibitem{Barlow1990}
R.~Barlow, {\it {Extended maximum likelihood}},  {\em Nuclear Instruments and
  Methods A} {\bf 297} (1990) 496.

\bibitem{Airapetian2009b}
{\bf HERMES} Collaboration, A.~Airapetian {\em et~al.}, {\it {Measurement of
  azimuthal asymmetries associated with deeply virtual Compton scattering on an
  unpolarized deuterium target}},  {\em Nucl. Phys. B} {\bf 829} (2010) 1,
  [\href{http://xxx.lanl.gov/abs/0911.0095}{{\tt arXiv:0911.0095}}].

\bibitem{Diehl1997}
M.~Diehl, T.~Gousset, B.~Pire, and J.~P. Ralston, {\it {Testing the handbag
  contribution to exclusive virtual Compton scattering}},  {\em Phys. Lett. B}
  {\bf 411} (1997) 193, [\href{http://xxx.lanl.gov/abs/hep-ph/9706344}{{\tt
  hep-ph/9706344}}].

\bibitem{Ji1998}
X.~Ji, {\it {Off-forward parton distributions}},  {\em Journal of Physics G:
  Nuclear and Particle Physics} {\bf 24} (1998) 1181,
  [\href{http://xxx.lanl.gov/abs/hep-ph/9807358}{{\tt hep-ph/9807358}}].

\bibitem{Belitsky2001}
A.~Belitsky, D.~M\"uller, L.~Niedermeier, and A.~Sch\"afer, {\it {Leading twist
  asymmetries in deeply virtual Compton scattering}},  {\em Nuclear Physics B}
  {\bf 593} (2001) 289, [\href{http://xxx.lanl.gov/abs/hep-ph/0004059}{{\tt
  hep-ph/0004059}}].

\bibitem{Goloskokov:2008ib}
S.~V. Goloskokov and P.~Kroll, {\it {The target asymmetry in hard vector-meson
  electroproduction and parton angular momenta}},  {\em Eur. Phys. J. C} {\bf
  59} (2009) 809, [\href{http://xxx.lanl.gov/abs/0809.4126}{{\tt
  arXiv:0809.4126}}].

\bibitem{Vandenbroucke}
A.~Vandenbroucke, {\em {Exclusive pi0 production at HERMES: Detection -
  simulation - analysis}}.
\newblock PhD thesis, University of Gent, 2007.
\newblock DESY-THESIS-2007-003.

\bibitem{Brasse1976}
F.~W. Brasse, W.~Flauger, J.~Gayler, S.~P. Goel, R.~Haidan, M.~Merkwitz, and
  H.~Wriedt, {\it {Parametrization of the $q^{2}$ dependence of $\gamma_{v}$p
  total cross sections in the resonance region}},  {\em Nuclear Physics B} {\bf
  110} (1976) 413.

\bibitem{MAID2000}
D.~Drechsel, O.~Hanstein, S.~S. Kamalov, and L.~Tiator, {\it {A unitary isobar
  model for pion photo- and electroproduction on the proton up to 1-GeV}},
  {\em Nucl. Phys. A} {\bf 645} (1999) 145,
  [\href{http://xxx.lanl.gov/abs/nucl-th/9807001}{{\tt nucl-th/9807001}}].

\bibitem{afanasev}
A.~V. Afanasev, M.~I. Kochatnij, and N.~P. Merekov, {\it {Single-spin
  asymmetries in the Bethe-Heitler process $e^{-}+p \rightarrow e^{-}+\gamma+p$
  induced by loop corrections}},  {\em Journal of Experimental and Theoretical
  Physics} {\bf 102} (2006) 220,
  [\href{http://xxx.lanl.gov/abs/hep-ph/0507059}{{\tt hep-ph/0507059}}].

\bibitem{Vanderhaegen2001}
M.~Guidal, M.~Vanderhaeghen, and P.~Guichon, {\it {Computer code for the
  calculation of DVCS and BH processes in the reaction ep $\rightarrow$
  ep$\gamma$}},  {\em Private Communication} (2007).

\bibitem{Chen2006}
{\bf CLAS} Collaboration, S.~Chen {\em et~al.}, {\it {Measurement of deeply
  virtual Compton scattering with a polarized proton target}},  {\em Phys. Rev.
  Lett.} {\bf 97} (2006) 072002,
  [\href{http://xxx.lanl.gov/abs/hep-ex/0605012}{{\tt hep-ex/0605012}}].

\bibitem{Musatov2000}
I.~V. Musatov and A.~V. Radyushkin, {\it {Evolution and models for skewed
  parton distributions}},  {\em Phys. Rev. D} {\bf 61} (2000) 074027,
  [\href{http://xxx.lanl.gov/abs/hep-ph/9905376}{{\tt hep-ph/9905376}}].

\end{thebibliography}

\providecommand{\href}[2]{#2}\begingroup\raggedright\endgroup

\end{document}